\documentclass[a4paper]{article}

\usepackage{INTERSPEECH2021}
\usepackage{hyperref}

\title{The Royalflush System for VoxCeleb Speaker Recognition Challenge 2022 }
\name{Jingguang Tian, Xinhui Hu, Xinkang Xu}
\address{
  Hithink RoyalFlush AI Research Institute, Zhejiang, China}
\email{\{tianjingguang, huxinhui, xuxinkang\}@myhexin.com}

\begin{document}

\maketitle
\begin{abstract}
In this technical report, we describe the Royalflush submissions for the VoxCeleb Speaker Recognition Challenge 2022 (VoxSRC-22). Our submissions contain track 1, which is for supervised speaker verification and track 3, which is for semi-supervised speaker verification. For track 1, we develop a powerful U-Net-based speaker embedding extractor with a symmetric architecture. The proposed system achieves 2.06\% in EER and 0.1293 in MinDCF on the validation set. Compared with the state-of-the-art ECAPA-TDNN, it obtains a relative improvement of 20.7\% in EER and 22.70\% in MinDCF. For track 3, we employ the joint training of source domain supervision and target domain self-supervision to get a speaker embedding extractor. The subsequent clustering process can obtain target domain pseudo-speaker labels. We adapt the speaker embedding extractor using all source and target domain data in a supervised manner, where it can fully leverage both domain information. Moreover, clustering and supervised domain adaptation can be repeated until the performance converges on the validation set. Our final submission is a fusion of 10 models and achieves 7.75\% EER and 0.3517 MinDCF on the validation set.
\end{abstract}

\section{System description for track 1}
Fully supervised speaker verification is a task where speaker labels are available when developing the system. In this track, we propose a powerful U-Net-based speaker embedding extractor with a symmetric architecture. We do experiments on the VoxSRC-22 dataset to demonstrate the superiority of the proposed network. Our final submission is the result of an individual system.

\subsection{Data}
\label{track1 data}
The experiments for the track1 are conducted on the Voxceleb2 development set, which contains 1,092,009 utterances from 5,994 speakers. The challenge released a brand new validation set that focuses on the impact of age from these speakers and shared background noise between speech segments from different speakers. We report experimental results on this new validation set which has 110,366 utterances and 306,432 trail pairs. 

Data augmentation increases the diversity and amount of training data from which neural networks can benefit. We generate 9 additional copies of the original training data as follows. The first four augmentations are generated using the publicly available MUSAN\cite{snyder2015musan} corpus(noise,music,babble) and the RIR(reverb) dataset\cite{ko2017study}. The remaining augmentations are created by sox. We use the tempo function to tempo down and up for each utterance from the Voxceleb2 development set by 0.9 and 1.1 factors without changing the pitch. In addition, we use the speed function to speed up and down each utterance of the original training dataset by 0.9 and 1.1 factors. The utterances with different speeds are considered new speakers. Finally, we have 1,092,009×9 = 9,828,081 utterances from 5,994×3 = 17,982 speakers.

\subsection{Speaker Embedding Extractor}
In our previous work, a U-Net-based network\cite{tian2022royalflush} was applied to detect overlapped speech, and it was found that such a network can also be applied to speaker verification. Detailed information about the network for this work is shown as follows. Three main components are contained:  representation layers in the frame level, temporal standard deviation pooling(TSDP) layer\cite{wang2021revisiting}, and a representation layer in the segment level.
The frame-level representation layers has three paths: the downsampling path, connection path and upsampling path. The downsampling path consists of three Squeeze-and-Excitation(SE) Conv blocks\cite{hu2018squeeze} whose feature maps are reduced to obtain a larger temporal context. The connection path contains 15 repeated Residual blocks\cite{he2016deep}. The upsampling and downsampling paths are symmetric and comprise three SE Conv blocks, where deconvolution is used to enlarge the feature map size. There are skip connections between the downsampling path and the upsampling path, and the corresponding feature maps are summed up as the input of the SE Conv block of the upsampling path. 
After frame-level feature learning layers, the TSDP layer, which computes the standard deviation of the output feature maps, can project the variable length input to the fixed-length representation. Finally, the 256-dimensional affine layer is adopted to extract speaker embedding. We named it ResUnet in the following report.
The detailed parameters of the architecture are shown in Table\ref{tab:table-U-Net}.

\begin{table}
  \caption{The ResUnet architecture, C(kernel size, stride) denotes the convolutional layer, the output size represents the channel, time, and frequency dimensions, respectively.}
  \label{tab:table-U-Net}
  \centering
  \begin{tabular}{ccc}
  \hline
  \textbf{Layer}     & \textbf{Structure} & \textbf{Output Size} \\ \hline
  SE Conv Block      & C(7 × 7, 1)        & 64 × 400 × 64        \\
  SE Conv Block      & C(3 × 3, 2)        & 128 × 200 × 32       \\
  SE Conv Block      & C(3 × 3, 2)        & 256 × 100 × 16       \\
  Residual Block × 15& C(3 × 3, 1)        & 256 × 100 × 16       \\
  SE Conv Block      & C(3 × 3, 1/2)      & 128 × 200 × 32       \\
  SE Conv Block      & C(3 × 3, 1/2)      & 64 × 400 × 64        \\
  SE Conv Block      & C(7 × 7, 1)        & 64 × 400 × 64        \\
  TSDP               & -                  & 4096                 \\
  Affine             & -                  & 256                  \\ \hline
\end{tabular}
\end{table}

\begin{figure*}[ht]
\centering
\includegraphics[scale=0.52]{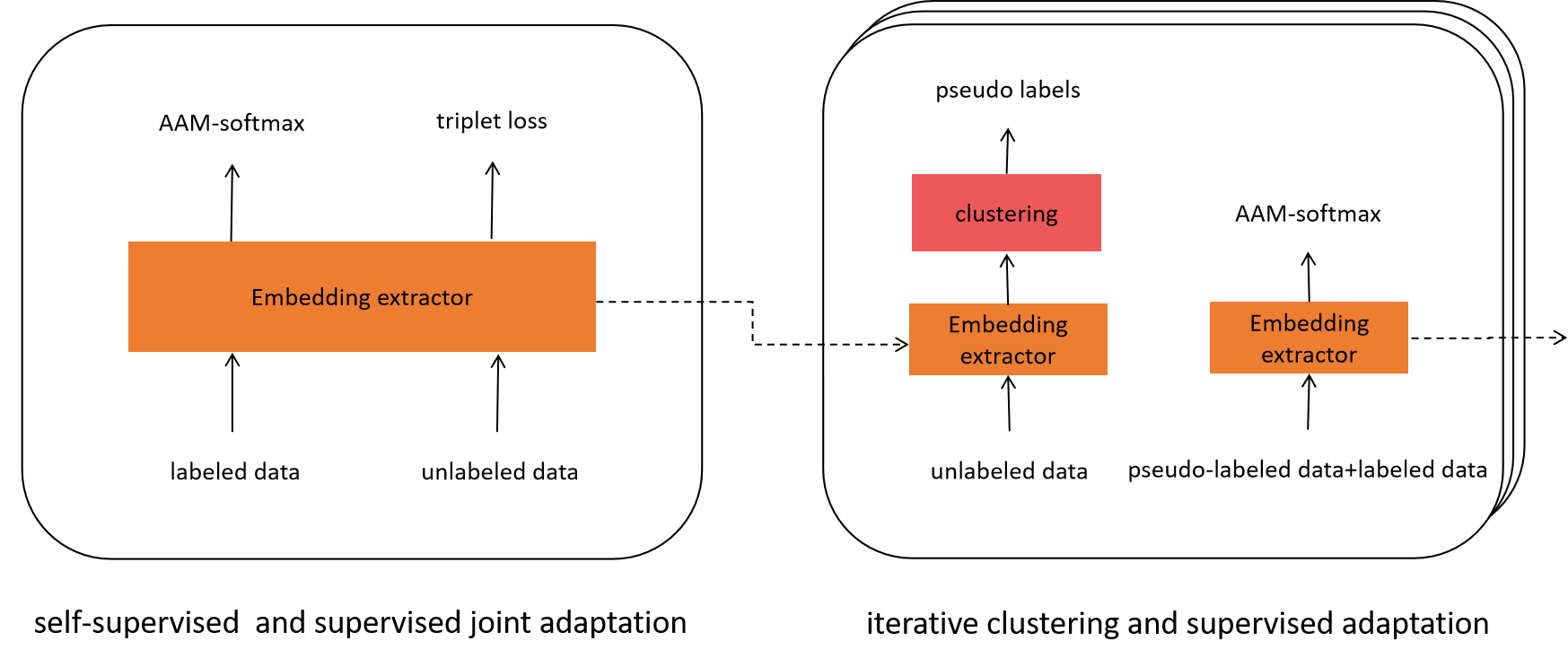}
\caption{The proposed two-stage domain adaptation framework for speaker verification}
\label{fig1}
\end{figure*}
\setlength{\belowcaptionskip}{-0.2cm} 

\subsection{Training Strategy}
\label{track1 training strategy}
The training method of the speaker embedding network has a large impact on performance. The input features are 64-dimensional log Mel-filterbank energies with a frame length of 25ms and a frame shift of 10ms. Utterance-level Cepstral Mean Normalization (CMN) by subtracting the cepstral mean from the whole utterance is applied to all input features. The training data is cut into 2-second length segments, and the mini-batch size for training is 128. The SGD optimizer with weight decay 2e-5 is employed. The speaker embedding extractor is trained using AAM-softmax\cite{deng2019arcface} with a scale of 32. We linearly increase the margin during the first two epochs and then fix the margin to 0.25. The initial learning rate is 0.1 and is multiplied by 0.9 after every 50,000 batches until the model converges. After the initial training, we apply a large margin fine-tuning strategy\cite{Thienpondt2021TheIV} to the model. During this fine-tuning stage, the margin of the AAM-softmax layer is increased to 0.35. The length of the random crop is increased to 4 seconds. The learning rate is fixed to 1e-4.

\subsection{Back-end Scoring}
The cosine similarity is the back-end scoring method. Neither score normalization nor score calibration is used. 

\subsection{Experimental Results}
To demonstrate the effectiveness of ResUnet, we compare it with ECAPA-TDNN\cite{desplanques2020ecapa}. The ECAPA-TDNN provides state-of-the-art speaker recognition performance through a careful architectural design based on recent trends in the face verification and computer vision. In our experiments, we set the number of feature channels of ECAPA-TDNN to 1024. The training strategy and data for the ECAPA-TDNN are the same as for the ResUnet. 

The performance of the developed systems are given in Table\ref{tab:track1}. The ECAPA-TDNN achives 2.60\% in EER and 0.1672 in MinDCF on validation set, and the ResUnet leads to a relative improvement of 20.7\% in EER and 22.70\% in MinDCF. Note that by fusing the scores of the two systems, there is no improvement in performance. We argue that it is due to the enormous performance gap between ECAPA-TDNN and ResUnet. Finally, we adopt the ResUnet results for final submission.

\begin{table}
    \caption{ EER and MinDCF performance of all systems on the track 1 validation set.}
    \label{tab:track1}
    \centering
    \begin{tabular}{ccc}
    \hline
        \textbf{Model} & \textbf{EER} & \textbf{MinDCF} \\ \hline
        ECAPA-TDNN & 2.60\% & 0.1672  \\ 
        ResUnet & \textbf{2.06\%} & \textbf{0.1293} \\ \hline
    \end{tabular}
\end{table}

\begin{table}
\caption{Details of training data presentation of track 3.}
\label{tab:data}
\centering
\begin{tabular}{ccc}
\hline
\textbf{Training Data}                 & \textbf{Utterances} & \textbf{speakers} \\ \hline
Labeled source domain data   & 9,828,081           & 17,982            \\
Labeled target domain data   & 9,000               & 150               \\
Unlabeled target domain data & 3,184,622           & ——                \\ \hline
\end{tabular}
\end{table}

\section{System description for track 3}
Semi-supervised domain adaptation for speaker verification aims to adapt a well-trained source speaker embedding extractor to target domain data from which unlabeled and limited labeled data are available. We propose a two-stage domain adaptation framework for this task. In the first stage, self-supervised and supervised learning work together to train the speaker embedding network. We can use this model to characterize the speaker in each unlabeled target domain utterance by extracting an embedding. In the second stage, these embeddings can be grouped through a clustering algorithm to generate pseudo labels that enable the speaker embedding network to perform domain adaptation in a supervised manner. We then use all source and target domain data in a supervised manner to adapt the speaker embedding extractor, which can fully utilize both domain information. Furthermore, clustering and supervised domain adaptation can be repeated with limited rounds. We illustrate the proposed framework in Figure\ref{fig1}.

\subsection{Data}
The challenge provides three training datasets. (1) a large set of labeled data in a source domain(Voxceleb2 development set), (2) a large set of unlabeled data in a target domain(a subset of the CnCeleb2 dev set which contains 454,946 utterances without speaker labels), and (3) a small set of labeled data in a target domain(a small set of CnCeleb data which has 1,000 utterances from 50 speakers). The official validation set involves 2,400 utterances from 120 speakers and 40,000 trail pairs.

Data augmentation is similar to section \ref{track1 data}. The unlabeled target domain data do not apply speed perturbation to increase speaker numbers. We list detailed statistics of the training data in Table\ref{tab:data}.

\subsection{Speaker Embedding Extractor}
We modify the number of residual blocks of our previously proposed speaker network, leaving other parts unchanged, to obtain 5 variants. The network in Table\ref{tab:table-U-Net} has 15 Residual blocks, we call it ResUnet-15. Follow the above naming rules, the other four are ResUnet-9, ResUnet-12, ResUnet-18, and ResUnet-21 respectively.

\subsection{Two-stage Domain Adaptation}
\subsubsection{Self-supervised and Supervised Joint Adaptation}
Self-supervised learning is a commonly used domain adaptation scheme. However, previous researchers have pointed out that using self-supervised learning only on the target domain dataset causes speaker embedding extractors to degrade speaker discrimination ability. To alleviate this problem, \cite{chen2021self} proposes a joint training strategy to exploit the rich information of the source and target datasets. We follow the approach described in subsection 2.3.2 of \cite{chen2021self}. Specifically, the labeled source domain data and unlabeled target domain data are simultaneously fed into the speaker network during the training phase. The output source domain embeddings are supervised by AAM-softmax loss, and target domain embeddings are supervised by triplet loss. At the same time, a small subset of labeled data in the target domain is also used to train the speaker network by AAM-softmax loss. Finally, these two losses are added together to optimize the speaker network. 

\subsubsection{Iterative Clustering and Supervised Adaptation}
In unsupervised speaker verification, iterative clustering methods have achieved impressive results\cite{thienpondt2020idlab}, and we employ a similar approach for domain adaptation. After a speaker embedding extractor is trained through self-supervised and supervised joint adaptation, we can use this model to extract an embedding from each unlabeled target domain utterance. An efficient minibatch k-means clustering is first used to reduce the number of embeddings, which can reduce the memory consumption of subsequent clustering. Then each utterance is assigned a speaker identity pseudo label by AHC. We treat these pseudo-labels as ground truths and then train a speaker embedding extractor by 2-subcenter AAM-softmax loss\cite{deng2020sub} using all source and target domain data. Furthermore, given the training embeddings generated by the model trained in a supervised manner, clustering and supervised training can be repeated until performance convergence on the validation set.

The number of clustering can be determined by analyzing the performance of the speaker embedding extractor on the validation set. In the first clustering step, we fix the number of clustering centers of mini-batch k-means to 20,000. In the second clustering step, we conduct several experiments, where the number of clusters ranges from 1,000 to 4,000, to find the optimal number of AHC. To ensure the stability of the training process, we filtered out pseudo-speakers with less than 10 utterances. The performance of ResUnet-9 trained by the 2-subcenter AAM-softmax loss with the different number of pseudo labels on the validation set is shown in Table\ref{tab:number of pseudo-labels}. The results demonstrated that 2,000 clusters delivered optimal results.

\begin{table}
\caption{The performance of ResUnet-9 trained with the different number of pseudo labels on validation set.}
\label{tab:number of pseudo-labels}
\centering
\begin{tabular}{cccc}
\hline
\textbf{Model} & \textbf{The number of clusters} & \textbf{EER}     & \textbf{MinDCF}  \\ \hline
ResUnet-9       & 1000                            & 11.28\%          & 0.52825          \\
ResUnet-9       & 2000                            & \textbf{10.90\%} & \textbf{0.47820} \\
ResUnet-9       & 3000                            & 11.00\%          & 0.49600          \\
ResUnet-9       & 4000                            & 11.34\%          & 0.49310          \\ \hline
\end{tabular}
\end{table}

\subsection{Training Strategy}
The training strategies are similar to section\ref{track1 training strategy} with a modification. We decrease the margin of sub-center AAM-softmax to 0.15 in the initial training and set the margin to 0.25 in the large margin fine-tuning stage. It is worth noting that during the second stage of domain adaptation, we fine-tune the model trained in the previous phase with a 1e-3 learning rate.

\subsection{Back-end Scoring}
We use cosine distance with sub-mean for scoring. The sub-mean is a simple trick to alleviate enroll-test mismatch. First, we extract the embeddings of all utterances in the validation set, then compute the mean embedding. The enrollment and test embeddings are normalized by subtracting the mean embedding before scoring.

We consider the duration of the utterance as quality measurement, and the duration-based score calibration\cite{Thienpondt2021TheIV} is adopted to make the output scores more robust in the case of varying duration conditions. We only consider the minimum duration value in enrollment and test, which ensures calibration is symmetric. To train calibration systems, we create a set of trials from the labeled target domain data. Note that each speaker verification system is individually calibrated using the same procedure.

\subsection{Experimental Results}
 We assume that models trained with different numbers of pseudo-speakers are complementary. The scores of the different systems are fused with equal weights in our experiments. The performance of the iterative clustering and supervised adaptation on validation set converged after 3 rounds. We list the results of all systems in the last two rounds in the Table\ref{tab:track3 performance}. 

The EER and minDCF metric are not consistent, some models have improved MinDCF but worse EER. ResUnet-21 achieves the best MinDCF of 0.3906 among all individual systems. ResUnet-15 obtains an EER of 8.52\%, which is the best of all individual systems. From round 2 to round 3, there is a slight increase in performance, which shows that it is saturated. The final submission system is a score-level fusion of 10 systems and achieves an EER of 7.75\% and a MinDCF of 0.3517 on the validation set. 

\begin{table}
\caption{EER and MinDCF performance of all systems on the track 3 validation set.}
\label{tab:track3 performance}
    \centering
    \begin{tabular}{cccc}
    \hline
        \textbf{Model} & \textbf{Iteration} & \textbf{EER} & \textbf{MinDCF}  \\ \hline
        ResUnet-9 & Round2 & 9.24\% & 0.4572  \\ 
        ResUnet-12 & Round2 & 9.18\% & 0.4410  \\ 
        ResUnet-15 & Round2 & 8.65\% & 0.4231  \\ 
        ResUnet-18 & Round2 & 8.69\% & 0.4089  \\ 
        ResUnet-21 & Round2 & 8.76\% & 0.3980  \\ 
        ResUnet-9 & Round3 & 9.07\% & 0.4449  \\ 
        ResUnet-12 & Round3 & 9.03\% & 0.4358  \\ 
        ResUnet-15 & Round3 & 8.52\% & 0.4107  \\ 
        ResUnet-18 & Round3 & 8.60\% & 0.3962  \\ 
        ResUnet-21 & Round3 & 8.70\% & 0.3906  \\ 
        Fusion of the all models & —— & \textbf{7.75}\% & \textbf{0.3517} \\ \hline
    \end{tabular}
\end{table}

\bibliographystyle{IEEEtran}

\bibliography{mybib}


\end{document}